# Quantum Communication Systems: Vision, Protocols, Applications, and Challenges

Syed Rakib Hasan[1], Mostafa Zaman Chowdhury[1,*], Md. Saiam[1], and Yeong Min Jang[2]

[1] Department of Electrical and Electronic Engineering, Khulna University of Engineering & Technology, Khulna 9203, Bangladesh

[2] Department of Electronics Engineering, Kookmin University, Seoul 02707, South Korea

Email Addresses: syedrakibhasanadit@gmail.com, mzceee@ieee.org, mdsaiam.eee@gmail.com, yjang@kookmin.ac.kr

**Abstract**

The growth of modern technological sectors have risen to such a spectacular level that the blessings of technology have spread to every corner of the world, even to remote corners. At present, technological development finds its basis in the theoretical foundation of classical physics in every field of scientific research, such as wireless communication, visible light communication, machine learning, and computing. The performance of the conventional communication systems is becoming almost saturated due to the usage of bits. The usage of quantum bits in communication technology has already surpassed the limits of existing technologies and revealed to us a new path in developing technological sectors. Implementation of quantum technology over existing system infrastructure not only provides better performance but also keeps the system secure and reliable. This technology is very promising for future communication systems. This review article describes the fundamentals of quantum communication, vision, design goals, information processing, and protocols. Besides, quantum communication architecture is also proposed here. This research included and explained the prospective applications of quantum technology over existing technological systems, along with the potential challenges of obtaining the goal.

## 1. Introduction

The way our uses of bandwidths are increasing in with it data rates, computational speed, and security are also increasing day by day. A massive number of bandwidths are provided through communication channels for every user in existing communication systems [1]. Internet of things (IoT), artificial intelligence (AI), virtual reality (VR), and many other technologies have developed so fast that these technologies have created a huge amount of traffic volume [2]. It can be expected that, between 2020 and 2030, every year worldwide mobile data traffic will increase by 55% [3]. By 2030, this increased traffic will generate 5,016 Exabytes [4]. The constant increase in wireless user devices, data uses, and the requirements for a higher level of user experience are the main factors influencing the advancement of the wireless network system quality of experience [5]. The current and upcoming wireless based networks will need to develop in several ways to meet these expectations and challenges. Recent technological components like long-term evolution and high-speed packet access are included as a part of the development of existing wireless based technologies [6]. The rapid increase in wireless broadband traffic has a profound influence on future mobile network architectures [7]. Many technologies are already being considered for a prospective wireless system through wireless research activities.

In the present day, we go through all possible techniques to overcome these challenges. However, every generation of technology has some limitations of its own. By overcoming those limitations, we stepped next to technology as a replacement for the previous one. What we see in these technologies around us is based on the classical theory of physics and the classical principles of physics. This classical theory or principle that we apply to establish new technologies, it also adds an extra dimension of limitations. To shatter these limitations, we find a promising and reliable solution from the quantum method that adds new dimensions over existing communication and computing technologies.

The need for a high computation power of the systems has risen swiftly along with the demand for communications that are quick, dependable, secure, intelligent, and environment friendly. It is apparent that quantum computing (QC) technology has the potential to surpass traditional computing systems because of the inherent parallelism provided by the core ideas of quantum mechanics and the prospects shown via recent breakthroughs [8]. In addition to service classification and autonomous management, machine learning (ML) is seen to offer great potential for handling the reconfigurability requirements of future systems. This quantum-powered computing and data-driven learning techniques have a greater chance of achieving the goals of a service-driven, fully intelligent sixth-generation (6G) communication network. Another emerging study field is QC-assisted communications, which is thought to have the potential for delivering exceptionally high data speeds and connection security in upcoming communication systems. This quantum communication technology is super fast, has high data rates, and is enormously secure. So, each person can have an uninterrupted and secure connection. In order to achieve these, research into the reliable communication of quantum channels for combined noiseless classical communication, quantum communication, and entanglement resources has started [9],[10]. This quantum communication also has clock synchronization and interactive communication with the brain computer.

Inspired by these new key features, we have taken a standardized approach aimed at a broad audience to comprehend the preliminaries, vision, architecture, information processing, advantages, and application scenarios of quantum communication. The contribution of this paper can be concise as follows :

- To help elucidate a clear overview of quantum communication, the primary concepts of quantum communication definition, quantum mechanics, and the difference between classical and quantum communication have been demonstrated.
- Ongoing research activities on quantum technology expounded to demonstrate the research track of quantum technology are presented.
- Discussion and overview of quantum protocols for better and deep understanding of information processing in quantum computing and communication.
- Presented quantum technology applications over existing technological infrastructure to get in-depth knowledge about the direction for future hybrid quantum technology.
- Discussed the challenges and future research challenges to accomplish the goal of quantum technology in the near future.

The rest of the paper is organized as follows. Section II presents preliminaries about quantum communication. A direction about the review methodology of this paper is provided in Section III. Section IV briefly explains the major protocols of quantum computing and communication technologies. An in-depth overview of quantum communication vision, design goals, proposed quantum communication architecture, and information processing is discussed in Section V. Section VI sets out the challenges and future research direction for accomplishing the quantum technology goal. At last, in Section VII, we reach our conclusions. Table I summarizes the numerous key acronyms used in this paper for ease of reference.

| | |
|---|---|
| 5G | Fifth Generation |
| 6G | Sixth Generation |
| AI | Artificial Intelligence |
| AR | Augmented Reality |
| COW | Coherent One Way |
| DPS | Differential Phase Shift |
| EPR | Einstein, Podolsky, and Rosen |
| FMF | Few Mode Fiber |
| FSO | Free Space Communication |
| ICT | Information and Communications Technology |
| IoT | Internet of Things |
| LEO | Low Earth Orbit |
| ML | Machine Learning |
| mmWave | Millimeter Wave |
| QBK | Quantum Bridge Keeper |
| Qubit | Quantum Bits |
| QC | Quantum Computing |
| QCiO | Quantum Computing-Inspired Optimization |
| QCN | Quantum Communication Network |
| QKA | Quantum Key Agreement |
| QKD | Quantum Key Distribution |
| Q-nodes | Quantum Nodes |
| QO | Quantum Optics |
| QSA | Quantum Search Algorithm |
| RF | Radio Frequency |
| SDN | Software Defined Network |
| SMF | Single Mode Fiber |
| SSP | Six State Protocol |
| THz | Tera Hertz |
| VLC | Visible Light Communication |
| VR | Virtual Reality |
| WAN | Wireless Area Network |

Table I. List of acronyms

## 2. Preliminaries

Quantum science brings a revolution to the world of physics. Applications of quantum science based on quantum principles and theories show a significant development over existing technologies. The communication system is not apart from these technologies. Quantum science in the communication sector has created a new dimension for future communication systems.

*2.1 What is Quantum Communication?*

The word quanta is the singular form of quantum. In physics, quantum is defined as a packet of energy. A unique particle called entangle photons, they share quantum states with each other. Quantum communication can be defined as the moving of quantum information as encoded in some quantum state from one place to another place. In this communication technology, data is not transmitted in bits. Rather, data is transmitted by quantum bits (qubits). In this information and communication technology (ICT), the whole system is end-to-end encrypted. Fig. 1 gives a clear overview of the basic secure quantum communication system. Here, the sender transmits an encrypted message to the receiver. The receiver can only open that message if the receiver has that secret key. Without having the secret key, no other person can access and open that message.

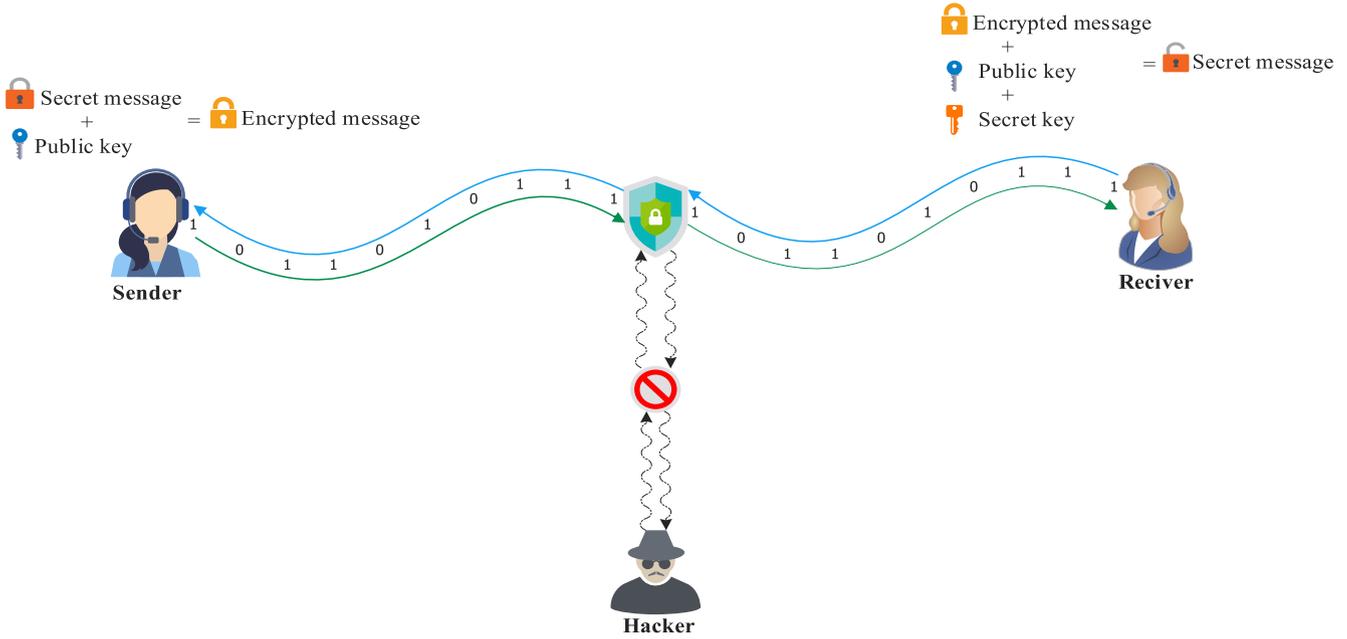

Fig. 1. Basic secure quantum communication.

## 2.2 Postulates of Quantum Mechanics

Quantum mechanics have six postulates, but four postulates rule the world of quantum communication and computing, which are the result of experimental observations. They share their place with axioms in the Euclidean geometry.

i. **First Postulate (State-space):** Any closed physical system's exact state may be characterized by a state vector $v$ with unit length in a Hilbert space $V$ and complex coefficients, that is an inner product with complex linear vector space (state space) [11].

ii. **Second Postulate (Evolution):** The development of any closed physical phenomenon in time may be described using unitary transformations that are dependent only on starting and the ending times of evolution [12].

iii. **Third Postulate (Measurement):** Any quantum measurement may be expressed in terms of a set of measurement operator $\{M_m\}$, where $m$ denotes the range of potential outcomes. $M_m$ is a matrix which is acting upon state space of the system that is being measured. $M_m^\dagger$ and $v^\dagger$ are the adjoints of the $M_m$ and $v$, respectively. The probability of measuring $m$ when the system is in state $v$ can be computed as [13],

$$P(m|v) = v^\dagger M_m^\dagger M_m v \qquad (1)$$

After measuring $m$, the system states can be defined as,

$$v' = \frac{M_m v}{\sqrt{v^\dagger M_m^\dagger M_m v}} \qquad (2)$$

Now considering the classical probability theory (1) can be expressed as,

$$\sum_m P(m|v) = \sum_m v^\dagger M_m^\dagger M_m v \equiv 1 \qquad (3)$$

Measurement operators must meet the completeness relation shown as,

$$\sum_m M_m^\dagger M_m \equiv I \qquad (4)$$

iv. **Fourth postulate (Composite system):** By applying the vector product of the separate systems, it is possible to calculate the state space of a composite physical system $W = V \otimes Y$. Moreover, system states $v$ and $y$ are defined corresponding to $y \in Y$ and $v \in V$ after that, the combined state of the composite system is $w = v \otimes y$ [14], where $v$ and $y$ are two dimensional vectors with unit length in a Hilbert space and complex coefficients.

Thus, the four basic postulates that rule the quantum realm of computation and communication.

## 2.3 Classical Communication versus Quantum Communication

In conventional communication systems, the transmitter transmits the sequence of 0 and 1 as data. Then, these sequence of 0 and 1 bits goes to the receiver by processing different steps. The main thing in the conventional communication system is that transmission is based on bits, however, in quantum communication, the transmitter transmits qubits. A single qubit can have the value of two bits. In general, $n$ qubits can have values of $2^n$ bits. Here, the main things are in the conventional system, if there is any branch created between receiver and transmitter, it can be detected, but in the quantum communication system, if there is any branch created between receiver and transmitter, it cannot be detected because of entanglement particles and quantum states.

## 3. Review Methodology

Selection criteria and research activities on quantum technology have been highlighted in the review methodology. In the selection criteria segment, the selection steps of this paper have been described and also a detailed overview of information selection in this paper is provided. In research activities on quantum technology part, some previous and ongoing research topics main focusing areas and contributions have been explained.

### 3.1 Selection Criteria

In the selection criteria section, the selection procedure of this paper is discussed. The review procedure, search criteria, database, exclusion, and inclusion criteria are all represented in this Table II. Two necessary steps are maintained to consist of this paper selection.

i. **Go through the title and the abstract in short:** At first, select the non-contradicting paper between 1976 to 2022. Then, read at a glance the paper's title and the abstracts to specifically point out the right paper. After selecting the proper papers, then jump into the next phase, which is full text go through.

ii. **Full paper goes through:** In this step, analyze the whole article based on the full text. After that, a decision was made on exclusion and inclusion for writing selection. The similar papers recently published in the journals and existed in the ArXiv were excluded. If more than one paper has been contributed by similar authors and their contributions are identical, then the most related paper is considered here.

Table II. Overview of selection topics

| Property | Category |
|---|---|
| Publication | Journal articles, international conferences, research articles, and scientific magazines. |
| Year consideration | 1976-2022 |
| Paper selection evaluation criterion | Initial overview paper title and the abstract. After that, select the proper paper and go through its full text. |
| Exclusion | Similar papers were recently published in the journals and existed in the ArXiv. |
| Inclusion | Focus on quantum communication working process, scenarios, protocols, applications, drawbacks, and new concepts |
| Paper selection classifications | Quantum communication protocol, quantum communication working process, challenges of quantum communications, vision of quantum communication, quantum communication application scenario. |

### 3.2 Research Activities on Quantum Technology

Different types of research, projects, and experiments are ongoing based on quantum methods around the world. In this part, some research activities are highlighted on quantum technologies over the past few years until now. Table III highlights the research direction, contribution, and the area of focus in some relevant publications. In other words, Table III is the summary of recent study on quantum technologies.

Table III. Summary of recent studies on quantum technology.

| No | Author | Reference | Year | Research direction and contribution | Area of main focus |
|---|---|---|---|---|---|
| 1 | T. Curcic et al. | [15] | 2004 | Different types of quantum communication applications, implementation, and quantum computing. | Quantum communication applications |
| 2 | S.T. Cheng et al. | [16] | 2005 | This paper proposed quantum routing mechanism that can teleport quantum sate wirelessly from one quantum device to another device. | Quantum routing mechanism for wireless network |
| 3 | P. Zoller et al. | [17] | 2005 | Quantum communication present status, goals, and visions. | Quantum information processing |
| 4 | N. Gisin et al. | [18] | 2007 | Different application areas of quantum physics, quantum universe, and approaches toward communications in the particle physics. | Quantum communication |
| 5 | Yu Xu-Tao | [19] | 2013 | It proposed distributed wireless quantum communication network by transferring the quantum state with the routing protocol. | Wireless quantum communication networks |
| 6 | G. Brennen et al. | [20] | 2015 | This paper mentioned about the necessity of quantum memory for long-distance communication and information processing. | Quantum memory |
| 7 | H. V. Nguyen et al. | [10] | 2016 | Communication through an asymmetric channel with a quantum coded realistic device. | Data transmission performance analysis via quantum coded device |
| 8 | S. Wehner et al. | [21] | 2018 | An overview of quantum internet, quantum internet development stages, the function of quantum repeater, and quantum internet application. | Quantum internet |
| 9 | S. J. Nawaz et al. | [22] | 2019 | An exhaustive review of machine learning, quantum computing, and quantum mechanics. Application of 5G network and quantum assist 6G communication is proposed. | Quantum computing and machine learning |
| 10 | A. S. Cacciapuoti et al. | [23] | 2019 | Differences between classical and quantum networks, transmitting information using quantum teleportation, and basic knowledge of quantum mechanics. | Quantum internet |

| No | Author | Reference | Year | Research direction and contribution | Area of main focus |
|----|--------|-----------|------|--------------------------------------|---------------------|
| 11 | R. N. Boyd et al. | [1] | 2019 | Harmless quantum cell phone implementation over existing RF frequency based communication technologies (4G and 5G). | Quantum cell phone |
| 12 | T. Huang et al. | [24] | 2019 | New architecture changes with 6G, different types of technology along with 6G, and new communication illustrations. | Green 6G communication and new paradigms of communication |
| 13 | M. Bhatia et al. | [25] | 2019 | Using quantum technology to analyze the performance of data transmission rate, task completion time, coverage area, and energy consumption | Quantum IoT based device performance analysis |
| 14 | D. D. Li et al. | [26] | 2019 | The design and implementation of the underwater quantum key distribution (QKD) based prototype and its performance measurement are presented. | QKD protocol for underwater communication |
| 15 | A. Manzalini | [27] | 2020 | Highlighted quantum optical communication application scenarios and technology trends. | Quantum based optical communication in ICT |
| 16 | A. Wallucks et al. | [28] | 2020 | Using quantum memory to store true quantum state and that can be useful for operation at highly efficient and low loss telecom wavelength. | Quantum memory |
| 17 | S. K. Singh | [29] | 2020 | Quantum computing blockchain diagram, quantum communication model system for future ICT and methodological flow. | Quantum communication for ICT |
| 18 | M. Bhatia et al. | [30] | 2020 | This paper proposed quantum computing focusing on network optimization in IoT device parameters to measure sensitivity, precision, and performance for quantum IoT applications. | Quantum IoT application |
| 19 | C. Ottaviani et al. | [31] | 2020 | Terahertz communication, performance analysis of THz quantum cryptography, and discussed physical hardware architecture of THz QKD scheme. | THz QKD based communication performance analysis |
| 20 | Chung et al. | [32] | 2021 | Quantum network architecture for optical fiber and controlling. | Quantum optical fiber network architecture |
| 21 | A. Kumar et al. | [33] | 2021 | This paper reviewed on different types of QKD protocols, challenges of quantum communication, and quantum cryptography. | QKD protocol |
| 22 | A. A. Abushgra | [34] | 2022 | This paper proposed different types of QKD protocols, improvements in cryptosystem, and security principles of QKD protocol. | Comparison between QKD and conventional protocol |
| | This paper | | | The vision, design goals, communication architecture, and protocols are presented. The prospective applications and challenges are discussed. | Overview of quantum communication |

## 4. Quantum Communication and Information Processing

Quantum communication is a recent technological development approach for modern communication systems. Therefore, vision and design goals must be set for this technology for further development. The vision for this communication system must be clear so that by keeps it in focus for further research activities continues. The quantum information processing and the proposed quantum architecture will give a clear concept of transmission to receiver end communication steps and the information processing steps between them.

### 4.1 Vision

The main visions of the quantum communication network are to maintain secure, stable, and high-speed communication for the different conditions of places and different type of users. This network not only supports our traditional data applications but also allows us to execute quantum services like QKD [35]. The future quantum device, quantum gadget, quantum computer, and all other devices can be connected to this quantum network. To make this network for daily life communication purposes, a relation bridge has to make with the existing network communication systems. For this reason, it is mandatory to develop a hybrid system of quantum classical network globally[35]. Fig. 2 shows a concept of future quantum communication scenarios. This technology spread influences over recent emerging technological development sectors such as fifth generation (5G) and beyond communication system, quantum computer, electronic banking, underwater communication, space communication, and machine-to-machine communication. The growth of technological development clearly imply that, very soon the bandwidth we are using won't be sufficient. To overcome this challenges quantum will add a new dimension to these technologies. As a result, high speed data communication is possible in 5G and beyond technological platforms. For high speed computing, quantum computing is a new approach. Today based on classical physics principles, the computation speed of the fastest supercomputers is nearly $10^{17}$ flops[36]. The computation speed of a quantum computer is much higher than any classical computer. Quantum computing and communication expand the roads of the banking sector. With the presence of quantum cryptography, banking transactions and security have become powerful. Different types of cryptocurrency have been used now for secure and anonymous transition. The beginning of a new era has opened by using quantum cryptography over blockchain technology. QKD based secure authentication blockchain protocol has the ability to maintain integrity and transparency of transactions against cyber-attacks [36]. Quantum computing has lifted the advancements of VR, augmented reality

(AR), and mixed reality. For having high computational processing power, quantum allows calculating high dimensional objects in low dimension [37]. As a result, it can solve very complicated pattern recognition problems in a short time. With AR, quantum makes it possible to create the most comprehensive mixed-media training applications, integrating VR with real-time action and telepresence from a distance for distant experts. QKD protocols are used in both free-space optical communication (FSO) for satellite to ground communication and underwater optical communication. Visible light communication (VLC) is compatible technology for indoor data communication. It allows better channel capacity than wireless fidelity. Using quantum in the VLC system will allow seamless, faster, and more secure communication [38]. Quantum key optimization method is a great approach for modern eHealth networks. By applying this technique, quantum bit leakage and transmission error were both significantly decreased [39]. As a result, it provides better medical diagnosis systems. Quantum cryptography in the eHealth sector also provides a secured network and keeps safe patient database records.

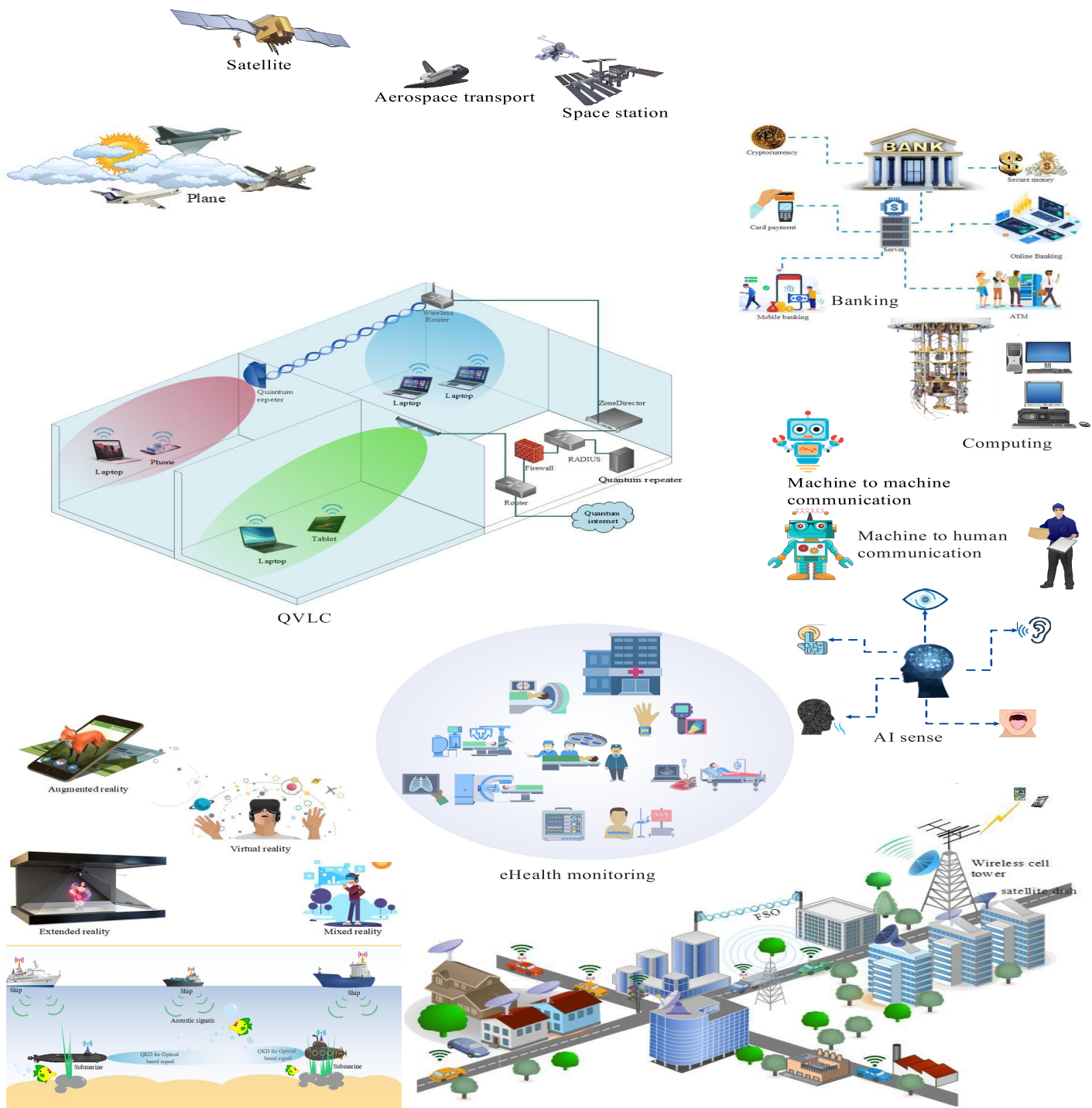

Fig. 2. Vision of quantum computing and communication.

*4.2 Design Goals*

Design is the most challenging issue for any compatible system of technology. If the design is not cost-friendly, simple, suitable, and sophisticated for the system, then the technology may not put any footsteps in the future. So, in this quantum communication network, keep all possible challenging things in mind to design it.

i. **Security:** The whole end-to-end communication in a quantum network must be highly secure. Otherwise, an invulnerable system may cause the hacking of important information.
ii. **Cost efficiency:** If this network is not cost effective, then it cannot be used for commercial purposes, and the highly funded project will be postponed someday for its high cost.
iii. **Shared infrastructure:** Shared infrastructure makes this network more suitable for multi-user communication.
iv. **Hybrid system:** Today, in world's existing communication technology is based on classical communication technology, and quantum communication is different from this classical technology. So, marching classical technology over quantum technology will provide greater benefits for both user and technological advancement.
v. **Manageability:** No matter what network it is, without proper management of this network, its efficiency and effectiveness will be poor.
vi. **High-speed stable connectivity:** Designing a network to minimize propagation delay and latency by ensuring a stable connection in any environment.
vii. **Simple hardware design:** Designing hardware for any technological device must be out of complexity. Then, its implementation will be ubiquitous, and its maintenance will be easy.

*4.3 Proposed Quantum Communication Architecture*

From transmission to the receiving end, the whole communication process needs to be end-to-end encrypted. The proposed communication architecture mainly focuses on end-to-end quantum channel, network management, QKD protocols, services, and applications. Fig. 3 shows the proposed quantum communication scheme from the transmission end to the receiver end. The transmission path is divided into two layers. One is the quantum physical layer, and another one is the quantum network layer. In the quantum physical node, two quantum nodes (Q-nodes) communicate with physical connectivity, signal rate, signal encryption, cryptography (secret key), generate random quantum numbers, and entanglement generation. In the quantum network layer, entanglement transmission and conventional wave routing are performed by software defined network (SDN) to establish Einstein, Podolsky, and Rosen (EPS) to Q-nodes or only between two Q-nodes. The Quantum channel is divided into two parts. The first one is the quantum network management and control layer, and the other one is the quantum link layer. Hardware and software management are the main functions of quantum network management and control layer. Link status monitoring and quantum nodes protocol layer are the key functions of the quantum link layer. Error control and flow control are a prime functions of link status monitoring. Quantum multidirectional protocol, secret sharing protocol, six state protocol (SSP), and SARG04 protocol are key protocols responsible for error control. On the other hand, coherent one way (COW) protocol, differential phase shift (DPS) protocol, and link layer protocol (LLP) are the primary protocols of flow control. Quantum nodes protocol layer build of BB84, BBM92, E91, S09, and S13 protocols. Finally, in the reception path, it is mainly divided into two layers. The first one is the quantum service layer, and the second one is the application. The quantum service layer is built on entanglement distribution and EPS management. At the end of all, confidentiality, authenticity, integrity, non-repudiation, and encrypted signal functions are under the application layer segment, which is the last part of the proposed quantum communication architecture.

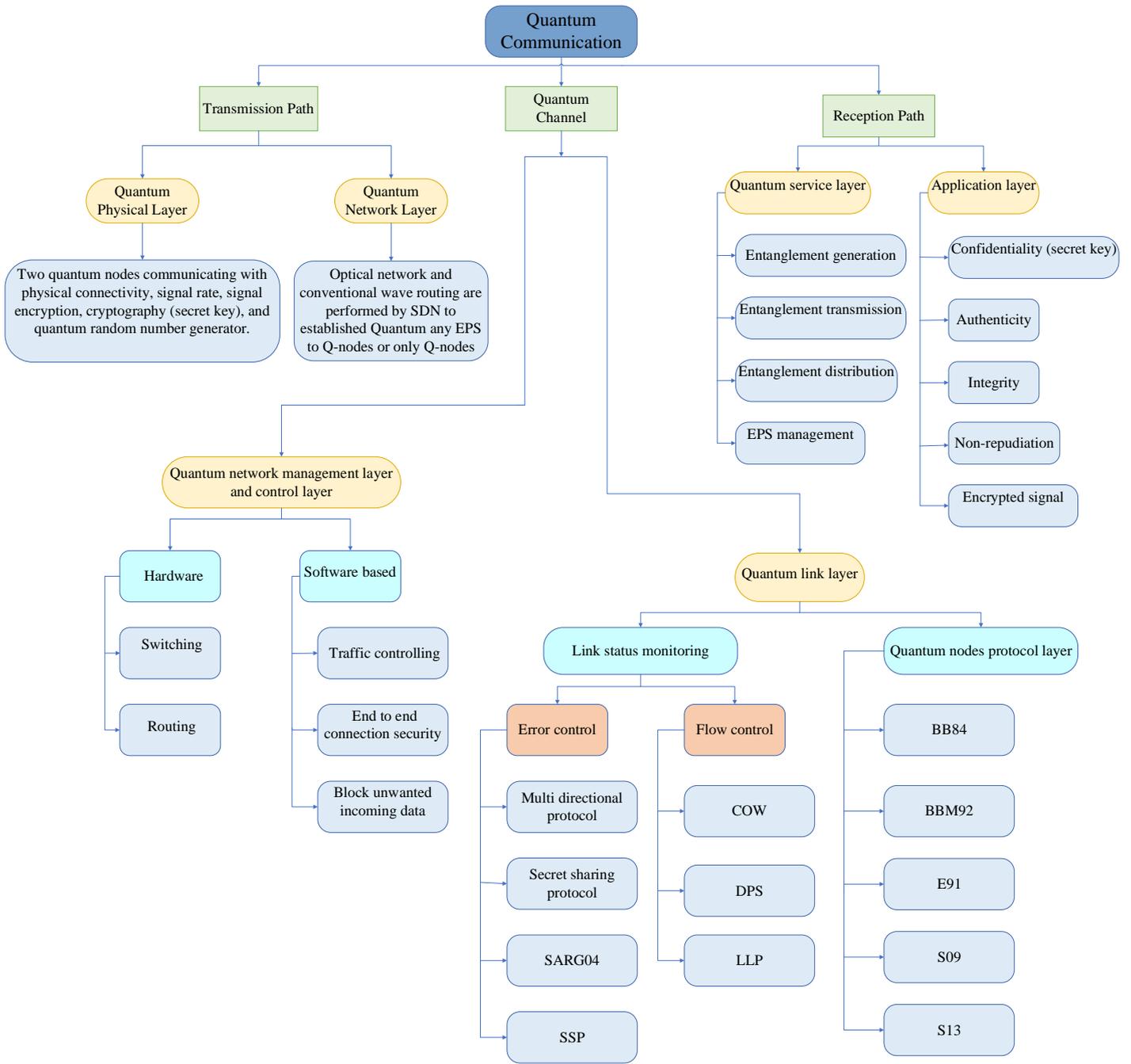

Fig. 3. Proposed quantum communication architecture.

*4.4 Quantum Information Processing*

The important elements of quantum information processing are quantum information science and quantum mechanism. They make it possible for quantum state transfer from one position to another. In the quantum communication process, two or more parts are connected with the quantum channel and classical communication channel for information processing. Usually, measurements are performed on the individual quantum subsystems, and the measurement bases used for every measurement are communicated via the classical channel. Here, this section concentrates on quantum communication along with discrete variables [40]. In quantum communication, it is subsisted on the parallel branch. To perform comprehensive theoretical and empirical works, it must be established on continuous variables that are the causes of the parallel branches.

There are seven stages in the development of the quantum information processing [40]. By completing each stage challenge with time, next stages are more complex than the previous stages. Fig. 4 shows the steps of quantum information processing. The first step is single physical qubit operation. Then, algorithm on multiple physical qubits. After that, in the third step control and error correction of qubits measure. In the next step, if the logical memory with a longer life time than physical memory, then it repeats first and second steps otherwise, it will allow fault tolerances and procced to compute for information processing. After completing all the challenges in every process then, quantum information processing establishes adequately.

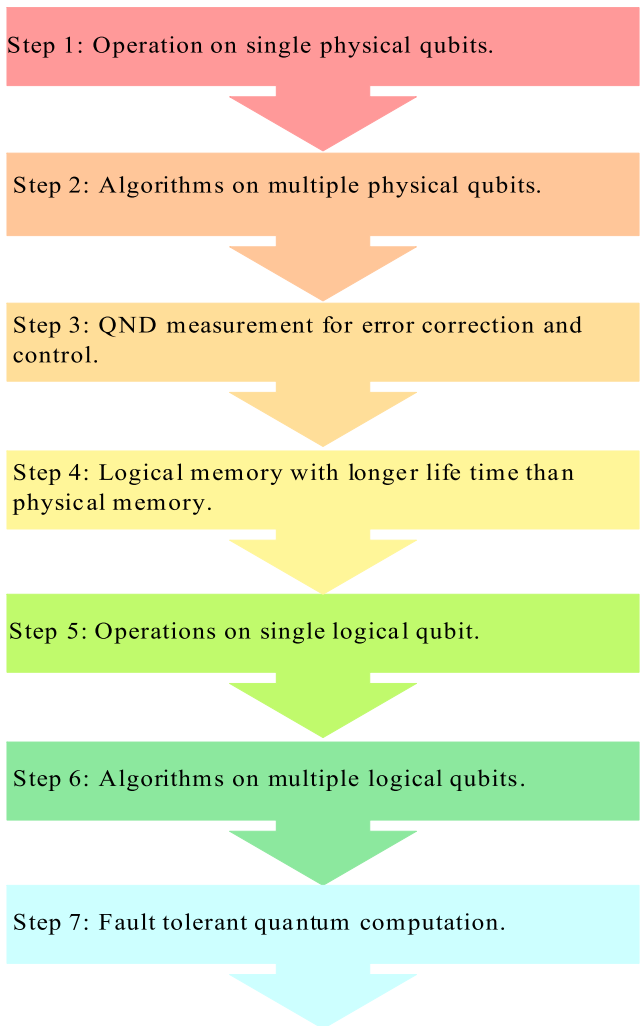

Fig. 4. Steps of quantum information processing.

**5. Quantum Key Distribution Protocols**

*5.1 Some major quantum key distribution protocols*

Cryptography is the process of transforming plain text data into incomprehensible text data in order to preserve information [29]. It is a process of storing and exchanging transaction data in a certain format. So that only those who are supposed to access and process it may do so [29]. The ability to distribute keys using quantum mechanics in a way that ensures Alice and Bob's security was one of the first breakthroughs in quantum computation and quantum information [41]. This method is referred to as quantum cryptography or QKD [41]. In previous communication technology, information was encoded from the transmitter side and decoded on the receiver side, this is the security or cryptography technology. Here the problem is if any third person wants to interrupt, then it is very hard to find the interrupter. In other words, it can be said that the process is not so secure. To make a secure connection, quantum cryptography brings a new solution. To ensure the high security of the system, the cryptographic protocol is must for the communication system. QKD protocols of quantum communication are used for secure quantum communication. Some major QKD protocols are briefly presented in this section.

*5.1.1 BB84 Protocol*

In 1984, QKD protocol was proposed for the first time by G. Brassard *et al.* [42]. This protocol is mainly designed based on Heisenberg's uncertainty principle and which is also familiarized as BB84 protocol [43].

The steps of BB84 protocol are given below [33]:

**Step 1:** Alice selects an *n* arbitrary bit using a flipping coin.

**Step 2:** Alice must flip the coin *n* times more to determine the basis, for each matching corresponding arbitrary bit.

**Step 3:** Bob receives the arbitrary bits that Alice prepared in their corresponding bases.

**Step 4:** Bob does not know the source corresponding to each arbitrary bit. Now, he tosses the coin *n* number of times. So that he can measures the received qubits in the obtained number of basis after tossing the coin. Bob declares the receiving of states.

**Step 5:** Using a classical channel, Alice and Bob compare their bases publicly. Alice tells Bob about the fundamentals of the agreement and the disagreement. Now, they drop the matching bit, if they disagree on a specific basis. After that, Bob will now get exactly n/2 random bits, and the remaining n/2 random bits are knocked out.

**Step 6:** Bob chooses half of the remaining n/2 arbitrary bits from Step 5 at random and compares them publicly with Alice. If they differ by more than the permissible number of mistakes (because to noise), they discard the entire series of arbitrary bits, indicating that Eve was listening. If these n/4 random bits are almost identical (i.e. within the permissible error because of noise). It means that, Eve was not listening. The remaining n/4 will be utilized as an arbitrary key between Alice and Bob in this circumstance.

*5.1.2 BBM92 Protocol*

For the first time BBM92 protocol was proposed by C.H. Bennett *et al.* [44]. In here, both Alice and Bob collect photons emitted from a single source and Alice is not supposed to produce a photon. Both classical information and quantum information are mixed in this protocol. The steps of the BBM92 protocol are given below:

**Step 1:** Alice generates strings of EPR pairs $q$ with the size $n$ and then transmits strings of qubits $q_b$ from every single EPR pair with the size of $n$ to Bob via a quantum channel. The other strings if qubits $q_a$ remain from every single pair with the size of $n$.

**Step 2:** Alice generates a string of bits with the size of $n$ arbitrary, which is denoted as $B_a$.

**Step 3:** Bob receives $q_b$ and then randomly generates string of bits $B_b$ with the size of $n$.

**Step 4:** Now, Alice measures every single qubit of $q_b$ corresponding to bits $B_a$ if $B_{a_i} = 0$, then it uses $x$ axis; else if $B_{a_i} = 1$, then it uses $z$ axis.

**Step 5:** After that, Bob measures every single qubit of $q_b$ corresponding to the bits of $B_b$ if $B_{b_i} = 0$, then it will use $x$ axis; else if $B_{b_i} = 1$, then it will use $z$ axis.

**Step 6:** Bob transmits his measurement axis by choices by $B_b$ to Alice via the public channel.

**Step 7:** After receiving $B_b$, Alice transmits her axis choice $B_a$ to Bob via the public channel and the Bob receives $B_a$.

**Step 8:** Alice and Bob both agree to remove all the instances in which they take place to measure different axis and instances in which measurements failed because detectors do not have perfect quantum efficiency. Then, the remaining of the instances can be used for generating private key $K_{a,b}$.

*5.1.3 Six State Protocol*

In 1999, for the first time, this SSP was proposed by H. B. Pasquinucci *et al.* [45]. By using three orthogonal bases SSP was designed by D. Baruß and he proved that its security is higher than BB84 protocol [46]. H. K. Lo has shown that if one way classical communication is allowed with SSP the bit error rate (BER) will be 12.7% which is a development over BB48 (11%) [47]. G. Kato *et al.* in [48] used a photon number for resolving detector to provide security proof of the SSP. Fig. 5 shows the indication of positive and negative of the Bloch sphere in *x, y,* and *z* axis, which are pointed by these six states.

The three bases of SSP are [33]:

Toward *z*-axis of the Bloch Sphere: $|0\rangle, |1\rangle$

Toward *x*-axis of the Bloch Sphere: $\frac{1}{\sqrt{2}}(|0\rangle + |1\rangle), \frac{1}{\sqrt{2}}(|0\rangle - |1\rangle)$

Toward *y*-axis of the Bloch Sphere: $\frac{1}{\sqrt{2}}(|0\rangle + i|1\rangle), \frac{1}{\sqrt{2}}(|0\rangle - i|1\rangle)$

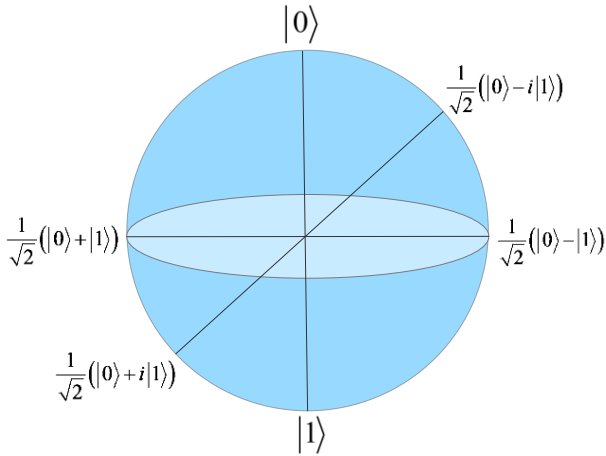

Fig. 5. The Bloch sphere, SSP basis states.

Alice chooses the basis with a chance of 1/3 and transmits qubits to Bob. If the input number is increased, then it is difficult for Eve to read the message. Alice uses a classical channel to announce the basis utilized, once Bob receives all qubits. The value which is used as key measured by Bob for Alice basis. Eve can use a random basis to measure the qubits transmitted by Alice and then transmit Bob new qubits. The probability of Eve gets the right basis is 1/3, and the probability that Eve chooses the wrong basis is 2/3. As a result, the probability of Bob getting the right qubits is 2/3.

*5.1.4 Differential Phase Shift Protocol*

DPS protocol was first introduced by K. Inoue *et al.* in 2003. Due to its relatively easy practical implementation, the DPS QKD algorithm is creating great interest. Coherent pulses are trained by this protocol that allows the particular users to protect from active beam splitting attacks [49]. The steps of DPS are given below :

**Step 1:** Alice generates a string of qubits $q$ with the size of $n$, which is carried by series of single photon, probably at four times consecutive intervals.

**Step 2:** Alice transmits $q$ to the Bob via the quantum channel.

**Step 3:** Bob receives $q$ by using detectors clicking at the time in the second and third interval, then records the time $T$ with the size $n$. Detector clicks into $D$ with the size $n$.

**Step 4:** Bob transmits $T$ to Alice via a public channel.

**Step 5:** Now, Alice receives $T$. From her modulation data and $T$, Alice knows which detector that click represents Bob's site.

**Step 6:** Then, Alice and Bob own uniform bit string, provided that the first detector clicks denotes "0" and the second detector represents "1". After that $K_{a,b}$ is formed. $K_{a,b}$ is denoted as shared raw key.

*5.1.5 Coherent One Way Protocol*

In 2004, N. Gisin *et al.* elaborated the COW protocol [50]. It is an improved QKD protocol designed to function with poor coherent pulses at high data rates [51]. This protocol's advantages include its experimental simplicity, tolerance of assaults, including photon numbers splitting and decreased interference visibility, and high efficiency in terms of exact secret bits per qubit. Keep in front of Fig. 6, the steps of COW are given below [52] :

**Step 1:** Alice generates strings of qubits $q$ with the size $n$ and $i$th qubit of $q$ is "0" with the probability $\frac{1-p}{2}$, "1" with the probability $\frac{1-p}{2}$ and the decoy sequence of the probability is $p$.

**Step 2:** Alice sends $q$ to Bob via the quantum channel.

**Step 3:** The items corresponding to a decoy sequence are sent by Alice as item A via a public channel.

**Step 4:** Bob removes all the detections at times $2A-1$ and $2A$ from his raw key. After that, he looks at whether the detector $D$ has ever fired at the time $2A$.

**Step 5:** Bob transmits Alice over the public channel $B$ of the periods $2A+1$, where he has a detector in $D$.

**Step 6:** When Alice receives $B$, she checks to see whether some of the items match the bit sequence "1, 0".

**Step 7:** Bob uses the public channel to transmit $C$, a list of the objects he found.

**Step 8**: Alice and Bob apply error correction and privacy amplification to these bits, and the private key $K_{a,b}$ is created.

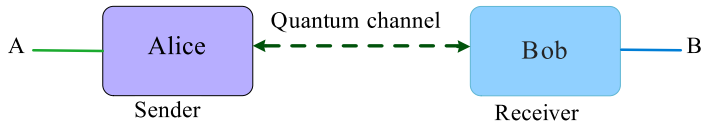

Fig. 6. COW protocol.

*5.1.6 E91 Protocol*

In 1991, E91 protocol was first proposed by A. Ekert. He tested eavesdropping by using generalized Bell's theorem. In order to generate identical random numbers at distant locations, his method utilized Bohm's version of the EPR. E91 is composed of the following steps [33]:

**Step 1:** Alice and Bob obtain their entangled photons from a prime source, in one out of four utmost entangled states ($|\psi_1\rangle$ to $|\psi_4\rangle$). Consider the source produce EPR pair $|\psi_3\rangle = |EPR\rangle = \frac{1}{\sqrt{2}}(|10\rangle + |01\rangle)$.

**Step 2:** Now, Alice measures her particle, which she received from entangle pair among any of the 0, 45, and 90 degree bases.

**Step 3:** Next, Bob measures his particle, which he received from entangle pair among any of the 45, 90, and 135 degree bases.

**Step 4:** After that, Alice and Bob measure the particle in a same base and remove the particle from a different base. If the base is the same, they obtain a common key. If any bits or particles remain that are known by shifted key.

**Step 5:** Finally, Alice and Bob check the system and then make the decision by interchanging the key's hashes, either accepting or throwaway the keys.

*5.1.7 SARG04 Protocol*

In 2004, V. Scarani *et al*. designed the SARG04 protocol vigorous against the photon number splitting attack [53]. C. Branciard *et al.* designed SARG04 protocol entangled version to prove that the performance of SARG04 is better than BB84 in terms of long distance communication, Eve's attack, and secret key rate [54]. From the SARG04 to the *n* sate protocol was generalized by M. Koashi [55]. This protocol also uses two non-orthogonal quantum states. SARG04 and BB84 both protocols have the same transmission and measurement phase. The first phases of SARG04 and BB84 protocols are the same. The main difference is in the second phase. In the second phase, Alice declares a pair of the non-orthogonal states [42]. Now, Alice and Bob double check which bases they have for which bits. Rather of stating her bases directly, she utilizes one of them to encode her bit. After that, if Bob chooses the correct base, he is able to measure the precise state; otherwise, he is not able to acquire the bit [53]. By using SARG04 protocol, H. K. Lo *et al.* successfully proved the security of photon pluses [43]. When Poissonian source produces a weak signal, an imprecise detector receives that signal SARG04 protocol is used in that situation [56].

*5.1.8 Quantum Secret Sharing Protocols*

For the first time, M. Hillery presented out a quantum secret sharing protocol built on the Greenberger-Horne-Zeilinger state that combines secret sharing with quantum technology to accomplish the goal of eavesdropping detection [57]. Its principle is based on quantum entanglement. Based on fundamental physics, quantum secret sharing protocol ensures unconditional security. A close connection exists between the security of this protocol and QKD [58]. Steps of quantum secret sharing protocol are given below [59]:

**Step1:** Alice initially creates three-particle entangled states, each of which takes the following form $|\psi\rangle = \frac{1}{\sqrt{2}}\left(|0\rangle \frac{|00\rangle + |11\rangle}{\sqrt{2}} + |1\rangle \frac{|01\rangle + |10\rangle}{\sqrt{2}}\right)$. After that, Alice sends the second and third particles of each entangled state to Bob and Charlie, respectively, and preserves the remaining particle for herself.

**Step 2:** When Bob receives a qubit, he chooses at randomly to determine and measure it using the basis $\{|0\rangle, |1\rangle\}$ and relays the state he discovered or to reflect it back to Alice without disruption. Charlie exhibits similar behavior.

**Step 3:** After temporarily restoring these qubits from Bob and Charlie, Alice declares that she has now obtained all the qubits over a public classical channel. Bob and Charlie then disclose the qubits they have measured.

**Step 4:** Depending on Bob and Charlie's selections, Alice chooses to do different operations on the qubits at her end. If Bob and Charlie both agree to measure the qubits on the basis $\{|0\rangle, |1\rangle\}$, then Alice measures her own qubit on the basis and uses the measurement result as a secret key bit. The following condition applies when $r_A$, $r_B$, and $r_C$ are denoted as the measurement results of Alice, Bob, and Charlie, respectively $r_A = r_B \oplus r_C$.

As a result, Bob and Charlie only can obtain Alice's secret key if they work together. Alice can utilize this decision to detect eavesdropping if Bob or Charlie decide to reflect their qubit.

*5.1.9 S09 Protocol*

In 2009, S09 quantum protocol is founded by E. H. Serna [60]. It is an arbitrary state protocol based on the Heisenberg uncertainty principle. The polarization of this protocol is bit-flip and phase-flip [61]. The efficiency and security of this protocol are average. Keep in front of Fig. 7, the basic steps of S09 protocol are given below [52]:

**Step 1:** Alice creates two strings of bits with the size *n* arbitrarily, which is denoted as $K_a$ and $B_a$.

**Step 2:** Alice creates a string of qubits *q* with *n* size, *q* is $|x_y\rangle$ in *i*th qubit, where *i*th bit of $B_a$ is *x* and *i*th bit of $K_a$ is *y*.

**Step 3:** Alice transmits *q* to Bob through the quantum channel.

**Step 4:** Bob receives *q* and after that arbitrarily generates a sets of bits $B_b$ with *n* size.

**Step 5:** Now, Bob measures every single qubit of $q$ correspondingly to a basis by bits of $B_b$. After completing the measurement, $q'$ is produced which is the elevation from of $q$.

**Step 6:** Bob transmits $q'$ to Alice via the quantum channel.

**Step 7:** Alice measures every single qubit of $q'$ to generate string $C$.

**Step 8:** Now, Alice sums $C_i \oplus B_{a_i}$ to achieve private key $K_{a,b}=B_b$.

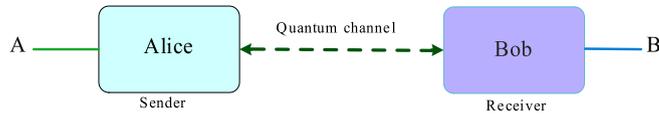

Fig. 7. S09 protocol.

*5.1.10 Link layer Protocol*

In quantum communication, the LLP was first proposed by A. Dahlberg in 2019. This protocol is robust for long distance communication.

The steps of the LLP are given below [62]:

**Step 1:** Alice randomly chooses $n$ number of strings from node A and then transmits it to node B. Bob receives qubits from node B.

**Step 2:** Node A and node B create one entangled pair. Then, node A sends qubits to node B correspondingly with respective qubits in the basis $r$.

**Step 3:** Node B sends the outcome strings to the node A.

**Step 4:** The error rates are estimated by both node A and node B.

**Step 5:** Now, node A and node B record their measurement results.

**Step 6:** After that, if Alice ensures the eavesdropping from node A, Bob shares key from node B.

*5.1.11 Key Agreement Protocol*

A significant area of study in cryptography is quantum key agreement protocol (QKA), which enables participants to agree on a conventional shared secret key across open quantum channels. In 2004, the QKA protocol was first proposed by K. F. Yu *et al*. Correctness, security, privacy and fairness these requirements should be met by a secure QKA protocol [63].

The steps of QKA protocol are given below, which is based on BB84 [64]:

**Step 1:** Any two $n$ numbers are selected by Alice randomly from bit strings $S_A = \{S_A^1, S_A^2, \ldots, S_A^n\}$ and $P_A = \{P^1, P^2, \ldots, P^n\}$. Where, $S_A^i$ and $P^i$ denote $i$th number of the bit for $S_A$ and $P_A$ respectively. Here, the range of $i$ is $1 \leq i \leq n$. Alice's contribution is represented by $S_A$ and $P_A$ denotes polarization bases that are utilized to encode $S_A$.

**Step 2:** As per the information of bit for each pair, Alice produces the associated photons into one of the four photon states that are listed below:

$$|\psi_{0,0}\rangle = |0\rangle = \frac{1}{\sqrt{2}}(|+\rangle + |-\rangle)$$

$$|\psi_{1,0}\rangle = |1\rangle = \frac{1}{\sqrt{2}}(|+\rangle - |-\rangle)$$

$$|\psi_{0,1}\rangle = |+\rangle = \frac{1}{\sqrt{2}}(|0\rangle + |1\rangle)$$

$$|\psi_{1,1}\rangle = |-\rangle = \frac{1}{\sqrt{2}}(|0\rangle - |1\rangle)$$

**Step 3:** After receiving protons, $n$ number of bits contributed $S_B = \{S_B^1, S_B^2, \ldots, S_B^n\}$. That is selected by Bobs to switch the photons states Alice received. After that, entire $n$ photons are saved into quantum memory by Bob.

**Step 4:** Bob ambiguously picks $n$ photons out of $m$ as public discussion to check, say it is $C$. Now, selected $m$ photons positions and an authorized classical channel connects $S_B$ to Alice is also declared by Bob.

**Step 5:** Following to the $S_B$ bits values, Alice determines $S_{AB} = S_A \oplus S_B = \{S_A^1 \oplus S_B^1, S_A^2 \oplus S_B^2, \ldots, S_A^n \oplus S_B^n\}$ as the key. After that, she transmits $P_A$ and constants of the $C$ (the values of $C$ are removed from $S_{AB}$ correspondingly position declared by Bob) to Bob by a classical channel.

**Step 6:** After receiving the message from the Alice, Bob measures all the photons correspondingly values of $P_A = \{P^1, P^2, \ldots, P^n\}$. Then the measurement result is decoded by Bob as key $S_{AB}$. He then compares the related measurement findings to obtain the $C$. If the comparing findings are identical, there are no eavesdroppers, and Bob will acknowledges Alice via a secure classical channel. After that, the Bob obtains the shared key $S = S_{AB} - C$. Which represents the leftover bits of $S_{AB}$ after eliminating the placements of the bits in $C$. Otherwise, Bob alerts Alice that the protocol is terminated if any error reaches the preset threshold.

**Step 7:** If Alice ensures that no eavesdropping occurs there, then she obtains $S = S_{AB} - C$ as shared key.

*5.1.12 KMB09 Protocol*

KMB09 protocol is also a QKD protocol [65]. The design goal of this protocol is to increase the transmission path without having any intermediate nodes [66]. Single photon source is required for this KMB09 protocol. Considering exceeding noise tolerance, long distance transmission can be done with this protocol. Robustness against eavesdropping is high in this protocol [65]. The steps of basic KMB09 protocol are given below [67]:

**Step 1:** Alice generates sets of strings with size *n* arbitrary, which is denoted by $K_a$. Then, each of the bit values is randomly assigned into an arbitrary index $i = 1, 2, 3, \ldots, N$ into $B_a$.

**Step 2:** Alice create sets of qubits *q* with *n* size, correspondingly either $|e_i\rangle$ or $|f_i\rangle$. Basis 'e' and 'f' is used accordingly here for coding '0' and '1'.

**Step 3:** Alice transmits *q* to Bob through the quantum channel.

**Step 4:** Alice transmits $B_a$ via a public channel.

**Step 5:** Now, Bob measures every single qubit of *q* by arbitrarily changing the basis of measurement among 'e' and 'f'. Then, he records the exact change into $K_b$, and the exact change information into the $B_b$.

**Step 6:** Bob transmits $B_b$ to Alice via public channel.

**Step 7:** Now, Alice and Bob determine the position of the bit where it remains. After that, rest of the bits of $K_a$ and $K_b$ are the private key of $K_{a,b}$.

*5.1.13 S13 Protocol*

In 2013, S13 quantum protocol was founded by E. H. Serna [68]. It is four state protocol based on Heisenberg uncertainty principle. The polarization of this protocol is two orthogonal. This protocol is the developed version of the BB84 protocol [69]. For the small distance communication, the security of this protocol is best, and efficiency is good. Keep in front of Fig. 8 the basic steps of S13 protocol are given below [52];

**Step 1:** Alice creates two strings of bits with the size *n* arbitrarily, which are denoted as $K_a$ and $B_a$.

**Step 2:** Alice creates a set of qubits *q* with *n* size, *q* is $|x_y\rangle$ in *i*th qubit, where *i*th bit of $B_a$ is *x* and *i*th bit of $K_a$ is *y*.

**Step 3:** Alice transmits *q* to the Bob through the quantum channel.

**Step 4:** Bob receives *q* and after that arbitrarily generates a string of bits $B_b$ with *n* size.

**Step 5:** Then, Bob measures every single qubit of *q* correspondingly to a basis by bits of $B_b$. Now, the result of measurement would be $K_b$, which has similar size of *n*.

**Step 6:** Alice transmits an arbitrary binary string, which is denoted as *C* to the Bob via a public channel.

**Step 7:** Alice sums $B_{a_i} \oplus C_i$ to achieve *T* and creates another arbitrary strings of binary value, which is denoted as *J*. *T* is set of bits which is generated by Bob. From all the elements that occupy a certain position *i* of the previous string, Alice will get new states of *q'* and then, transmits it to Bob via quantum channel.

**Step 8:** Bob sums $1 \oplus B_{b_i}$ to get the binary basis string *N* and measures *q'* following to these bases, and then generates detector *D*.

**Step 9:** Alice sums $K_{a_i} \oplus J_i$ to get binary string *Y* and then transmits it to Bob via a public channel.

**Step 10:** Now, Bob encrypts $B_b$ to get *U* and transmits to Alice via a public channel.

**Step 11:** After that, Alice decrypts *U* to get $B_b$. Then, she sums $B_{a_i} \oplus B_{b_i}$ to get *L* and transmits *L* to Bob via a public channel.

**Step 12:** Bob sums $B_{b_i} \oplus L_i$ to get $K_{a,b}$, which is a private key.

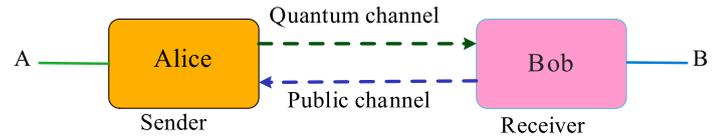

Fig. 8. S13 protocol.

*5.2 Observation of Protocols*

In a network system, the protocol is a series of rules for transmitting, receiving, and processing data. In Section IV, a brief overview of major QKD protocols are discussed. Table IV summarizes the protocols overview in short, which is elaborately explained above.

Table IV. List of protocols, principles, and polarization

| No | Name of protocol | Founder | Reference | Year | Principle | Polarization |
|---|---|---|---|---|---|---|
| 1 | BB84 | C.H.Bennett *et al.* | [33], [66], [70] | 1984 | Heisenberg uncertainty | 2-orthogonal |
| 2 | E91 | A. Ekert | [71], [72] | 1991 | Quantum entanglement | Random |
| 3 | BBM92 | C.H. Bennett *et al.* | [44], [73] | 1992 | Heisenberg uncertainty | Non-orthogonal |
| 4 | SSP | H. B. Pasquinucci *et al.* | [46], [74] | 1999 | Heisenberg uncertainty | 2-orthogonal |
| 5 | Secret Sharing | M. Hillery *et al.* | [75], [76] | 1999 | Quantum entanglement | 2-non-orthogonal |
| 6 | DPS | K.Inoue *et al.* | [34], [77], [78] | 2003 | Quantum entanglement | 4-non-orthogonal |
| 7 | SARG04 | V. Scarani *et al.* | [53], [66] | 2004 | Heisenberg uncertainty | 2-non-orthogonal |

| No | Name of protocol | Founder | Reference | Year | Principle | Polarization |
|---|---|---|---|---|---|---|
| 8 | Key Agreement | K. F. Yu *et al.* | [79], [80] | 2004 | Public private key cryptography | 2-orthogonal |
| 9 | COW | N. Gisin *et al.* | [33], [50] | 2004 | Quantum entanglement | non-orthogonal |
| 10 | KMB09 | M. M. Khan *et al.* | [34], [65] | 2009 | Heisenberg uncertainty | Orthogonal or non-orthogonal |
| 11 | S09 | E. H. Serna | [61], [81] | 2009 | Public private key cryptography | Bit-flip or phase-flip |
| 12 | S13 | E. H. Serna | [61], [82] | 2013 | Heisenberg uncertainty | 2-orthogonal |
| 13 | LLP | A. Dahlberg *et al.* | [62] | 2019 | Quantum entanglement | Undefined |

## 6. Application Scenarios

Many experiments, research, and projects are going on over quantum technological developments. As a result, many technological fields have been created for the application of quantum technologies. Some major applications are discussed below,

### 6.1 Quantum Optical Communication

A research and development path addressing the phenomena (and associated technologies) relating to the interactions between quanta of the electromagnetic field and matter is known as quantum optics (QO) [83]. Initial research on QO concentrated on several straightforward non-classical light states such as single photon, compressed states, twin optical beams, and EPR states; which are only defined by a few electromagnetic field modes [83]. By using quantum optics theory and principle, there will be the establishment of quantum optical communication. The idea behind quantum optical communications is to use photons, which are electromagnetic field quanta, as flying qubits that can transmit qubits from a physical quantum emitter to a physical quantum receiver across a network [27]. The benefits of flying qubits are low environmental effects, low noise, and high emission speed over the radio and optical channels. Use of quantum cryptography, this network is highly secure. We know that terahertz (THz) system can be operated by QKD. As a result, the architecture of the physical hardware of THz technology can be used in QKD technology. For this reason in 6G and upcoming new network technologies, the development and implementation of quantum optical communication is a new and promising approach.

### 6.2 Cluster States Based Quantum Communication Network with the Help of FSO

A cluster-state computation starts with the concoction of a special entangled many qubit quantum states known as a cluster state, which is then followed by an adaptive sequence of single-qubit measurements that processes the cluster, and finally, the read-out of the computation's result from the remaining qubits [84]. Cluster state refers family of quantum states. To allow the next generation of quantum communication networking, a situation is proposed recently, in which unconnected terrestrial cluster state based quantum communication network (QCNs) are coupled through the low earth orbit (LEO) satellite (cluster state) quantum network to enable worldwide coverage [85]. The two-dimensional cluster state is ubiquitous, for this reason, this same network architecture may be used for both QCN and distributed QC [85]. Considering the case when each node has several qubits and numerous layers of two dimensional cluster states are active at once, allowing us to execute QCN and distribute quantum computing at the same time [85]. With the help of quantized detector networks, it can able to configure QCN. The function of software defined network (SDN) is to ensure service in network management, data plane, and control plane. SDN based QCN architecture is divided into three layers: the application layer, the control layer, and the QCN layer [85]. In application layer request has been sent through the user. For this process control layer help him. Dense wavelength-division multiplexing/FSO/single-mode fiber (SMF)/few-mode fiber (FMF) and other nodes of QCN are compressed in QCN layer. Two nodes into QCN can be communicated by FSO/FMF/SMF link or a channel of wavelength which is dedicated to it.

### 6.3 Quantum Based Satellite Communication

Both classical and quantum information can be transmitted through the quantum channel. Fig. 4 shows the basic scenarios of quantum based satellite communication. Classical information is transmitted by the sender (plane, ships, submarines, base station) to the receiver through a satellite quantum channel. The whole converting process of communication begins from the classical domain to the quantum domain and is done by classical input and encoding. In the classical state, this is equivalent to source coding and channel coding [86]. And by the channel, data is transmitted. Because of the channel errors, receiver receives some error bits. Now it decodes quantum transformation and measures data.

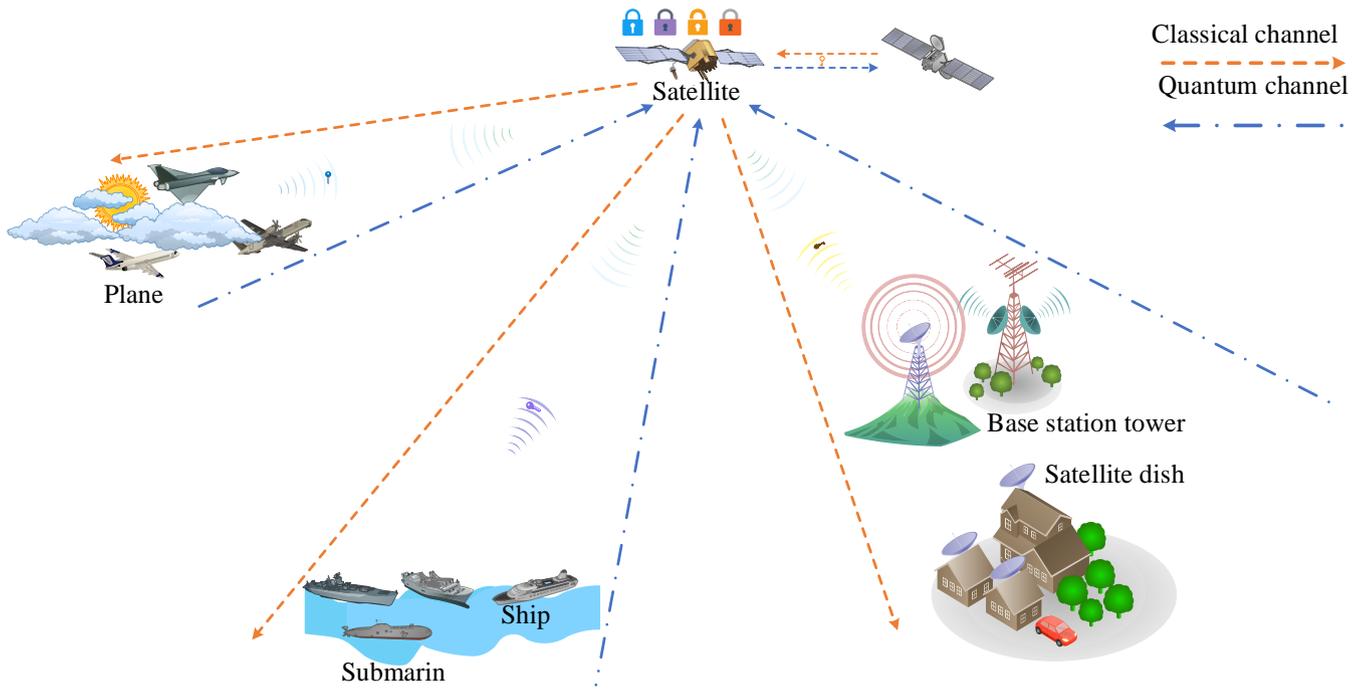

Fig. 4. Quantum based satellite communication.

*6.4 Terahertz Quantum Communication*

THz means trillion cycles per second. THz waves lie between the long wavelength of the infrared and short wavelength of microwave radiation. The term "THz waves," also known as "submillimeter radiation," typically refers to the frequency range between 0.1-10 THz, with corresponding wavelengths in the 0.03-3 mm range [87]. Using the range of this wavelength for communication purposes is called THz communication. Though THz communication is high-speed communication but, the ranges of this communication system security is so low between one meter to one kilometer distance. For long transmission and without compromising its security, the approach has been taken as THz quantum cryptography. In here, the communication method is the same based on THz and the security is ensured by quantum cryptography. This high-frequency signal is encoded by quantum technology and transmitted through a classical channel across the quantum channel to decode this information. Here secret key is also needed. To implement quantum cryptography in THz communication, special hardware must have to be implemented [31].

*6.5 Quantum Teleportation*

The application of quantum entanglement is quantum teleportation. Quantum teleportation is the method by which the state of one quantum system is transmitted to another distant quantum system without the need for any intermediate site [88]. It is the quantum information, or rather the quantum state that is transferred and not the physical system on which the qubit state is stored [12]. A complete quantum teleportation process, from start to end, goes through three steps. The first step is distribution of entanglement, the second step is Bell-state measurement. The last step is the sender informing the receiver of the outcome of this Bell-state measurement. Completing these operations teleportation process is finished.

*6.6 Quantum Communication in 5G and 6G*

The way our network users is increasing with this demand also rising proportionally. Our aim is not only fulfilling this demand, we must ensure the system of the network's security, reliability, and capability. To meet these challenges, we stepped into new 5G and beyond technologies. These technologies have some extra features such as AI, AI based networks, global connectivity, and sense. Implement the quantum communication technology in 5G and beyond networks the data rate will increase, link capacity will be increased and almost noiseless communication will be achieved. This technology comes with low latency, high mobility (up to 1000 km/hr), and high operating frequency (up to 1 THz). So it can be said that this network has a high security and data rate. In future communication techniques, that are being developed to increase link capacity, such as power domain multiple access supported by successive interference cancellation, have very high run-time computational power requirements. Therefore, there is a clear opportunity to utilize quantum communication [22]. The intrinsic security characteristic of quantum entanglement, which cannot be replicated or accessed without tampering, makes it an appropriate technology for 5G and beyond networks [4]. Fig. 9 shows some implementations of QKD, and other protocols work properly on this system. QKD based optical communication added a new dimension for FSO. Quantum communication also allows high security in autonomous vehicle communication. As a result, any third person cannot track the vehicle information during the journey time. This quantum based secured vehicle communication is given high priority for very important person transportation. QKD based secured wireless communication allows to block any third person, who wants track or crack the information between sender and receiver. Another emerging study topic is QC-assisted communications, which is expected to show promise for obtaining exceptionally high data speeds and connection security in future 6G and beyond communications [22]. In wireless communication networks, Quantum communication also solves different types of problems in numerous layers such as minimizing the latency, maximizing the transmission data rate, and providing better security. So, it can be seen that quantum communication technology can make a vital impact on 6G communication.

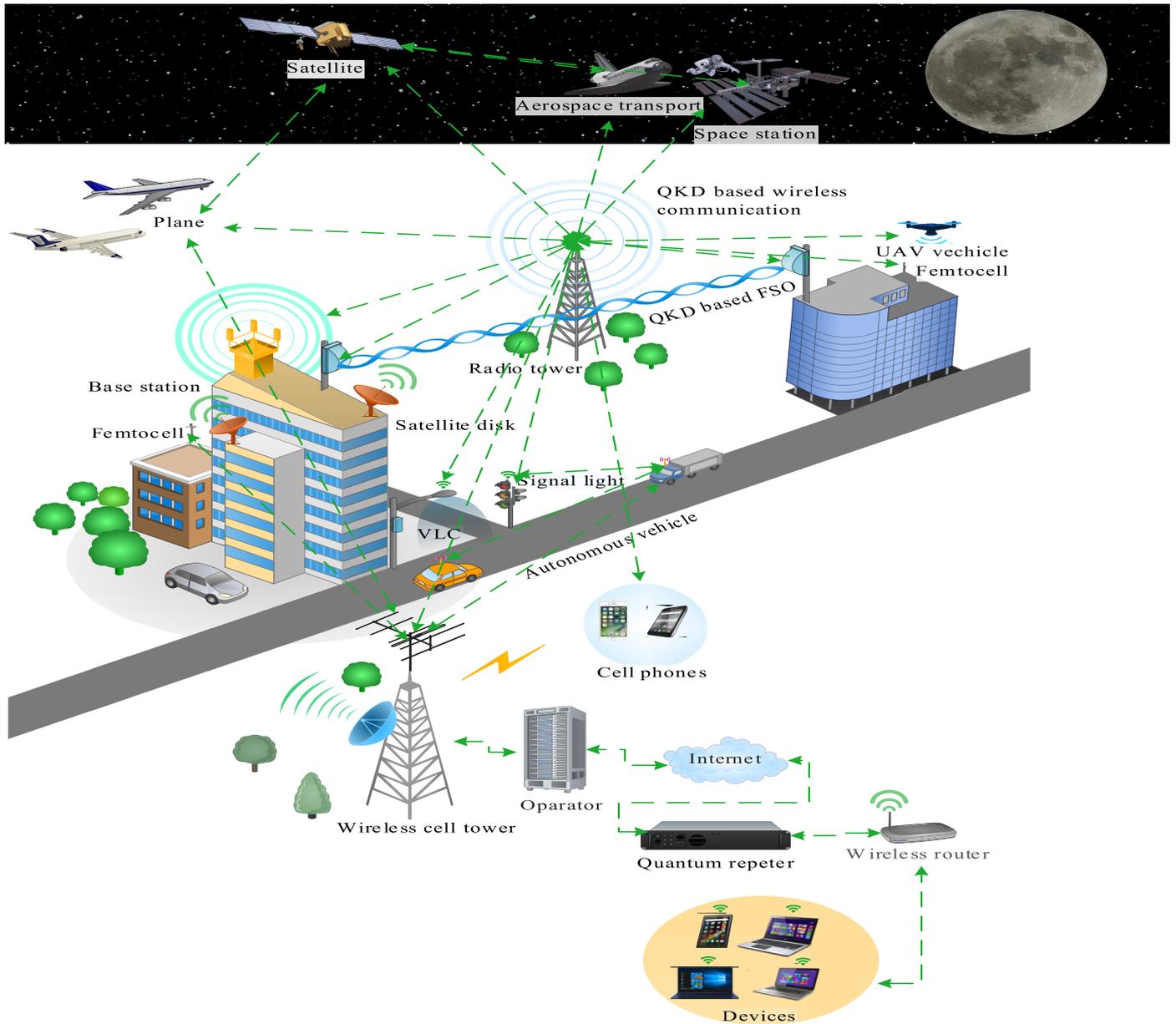

Fig. 9. Quantum communication in 5G and 6G communication scenarios.

*6.7 Quantum Assist Indoor Localization for mmWave and VLC*

Millimeter-wave(mm-Wave) and VLC technology are the latest technologies for indoor localization applications. One of the main reasons for implementing these technologies is to get the proper location with high accuracy. The previous methods of localization basis on fingerprints which is based on radio frequency (RF) and VLC technologies. However, their accuracy of localization is both complex and costly. So, one of the promising approaches to address this complexity reduction issue is to employ a quantum search algorithm (QSA), which aims to find the minimum entry in the unsorted database with *N* elements by utilizing some techniques of cost function evaluations. The possibility of using a QSA is to reduce the computational complexity of mm-Wave and VLC based localization techniques while attaining the same performance as a comprehensive search [22].

*6.8 Quantum Computing*

Quantum computing is a branch of research that employs quantum mechanical phenomena such as superposition and entanglement to conduct operations on data [14]. To overcome the limitations of classical computer, quantum computer is a great approach. Different types of problems and algorithms are solved by quantum computer so quickly than the classical computer. The algorithms used here are so complex. There are some challenges

to meet for the establishment of quantum computers. The challenges of quantum computer are

i. The amounts of qubits must be increased.
ii. Qubits must have some random values which can be initialized.
iii. Quantum gates must be faster than the loss of the quantum coherence time.
iv. It must be some unique set of universal gates.
v. Easily readable qubits.

*6.9 Quantum Memory*

Quantum memory is one of the major parts of quantum communication systems. For long-distance communication without missing any information, quantum memory is a great approach. It is now possible to increase the range of quantum communication with the help of quantum memory. The operation and synchronization of quantum memory are extremely time consuming [86]. Quantum memories are used to convert qubits between light and matter in order to achieve the scalability necessary for large-scale and long-distance quantum communication [86].

*6.10 Quantum Aided Multi-user Transmission*

Multi-user transmission means many information or data are sent through multipath of one system to another system. Quantum-aided multi-user transmission means that multi-user transmission will happen with the help of quantum technology. QSA can be used to reduce the complexity of vector perturbation precoding and improve multi-user transmission performance in wireless networks [22]. For this reason, multi-user transmission is now done smoothly. In order to execute vector perturbation precoding, reduce transmission power at the base station, and lower BER at mobile users, quantum-assisted particle swarm optimization methods can be used in both discrete and continuous modes [22]. By so many experiments conducted on it, the result shows that with the help of quantum precoding bit error rate is minimized and gets better performance.

*6.11 Quantum Internet*

If the transaction of information in a network through quantum devices are based on the principle of quantum mechanics, then this network is called as quantum internet [21]. Data can be encoded in the state of qubits, which can be created in quantum devices like a quantum computer or a quantum processor. Qubits can be sent through a network of numerous quantum devices. These devices are separated by physically. There are mainly six major functionalities for quantum internet. They major functionalities of quantum internet are,

i. Quantum teleportation.
ii. Quantum channel.
iii. Quantum repeaters.
iv. Quantum memories.
v. QKD.
vi. End node.

*6.12 Underwater Quantum Communication*

Over 70% of the earth's surface is covered by oceans. Underwater oil exploitation, oceanographic investigations, and underwater military actions are just a few instances of how the waters are being explored for industrial, scientific, and military interests [89]. If quantum technology can be added to underwater communication, then it will be a revolutionary change for underwater communication technology because water covers a major part of the earth's surface. Due to a variety of physical processes in varied underwater settings, from shallow coastal water to deep sea or oceans, dependable underwater communication faces substantial challenges [90]. To get rid of these interruptions to underwater communication quantum method is a great approach. Fig. 10 shows underwater QKD based communication. It explains the QKD based RF communication among satellites, ships, submarines, and QKD based optical communication between submarines. The major motive for exploring optical based QKD in an underwater setting is that, present acoustic communication technology has limited bandwidth and cannot be communicated over the water systems [91]. The average polarization fidelity for the optical based QKD in an underwater communication of the whole connection
is greater than 98% [26]. Using of QKD protocol for underwater communication will be more secure, and it would be possible to minimize the noise due to underwater turbulence by using quantum communication.

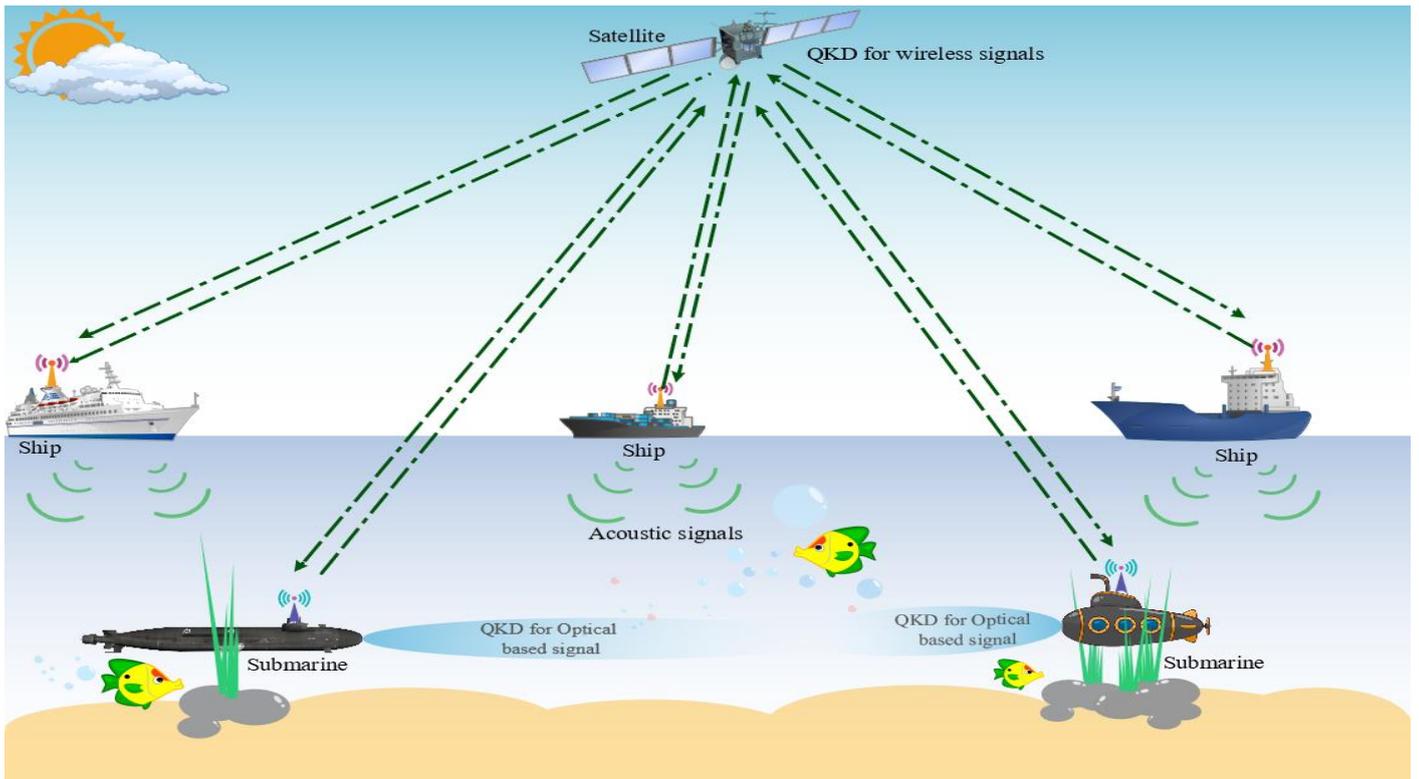

Fig. 10. Underwater QKD based communication.

*6.13 Quantum IoT*

IoT has evolved steadily over the last decade [92]. The advancement of IoT technology has laid the groundwork for innovative applications in a wide range of variety in industrial domains such as agriculture, logistics, and healthcare [93]. The applications of IoT are growing so faster over a few decades. IoT working system has four fundamental principles. They are given below,
  i. Sensors.
  ii. Connectivity.
  iii. Data processing.
  iv. User interference.

Combining these four fundamentals, IoT system works. To make this technology as a ubiquitous communication purpose there are lots of new things and technologies must be added with this system. To take this technology into a new level, we must make sure its security, stable connectivity, and network strength. Quantum technology will give this IoT platform a new level. Quantum computing improves the computational and data processing capability of IoT sensors and devices [25]. IoT sensor space minimization is a dynamic challenge inspired by the traditional sensor placement problems [30]. Various optimization approaches have been used for this goal; however, recent advances in quantum computing-inspired optimization (QCiO) have opened up new paths for obtaining the optimal behavior of the network system [30]. Fig. 11 shows QKD based QCiO IoT network system. In this network system a user first gives an input command to the device. Then, the command goes to the server. After that, by using QKD protocol the server processes and optimizes the data. This data optimization is done with the help of QCiO. Now, data reaches the gateway. Gateway optimizes the network to ensure the best routing path in the network and makes secure communication for example in the industry, robotics, smart home, banking, IoT based traffic signal, and power generation systems. This QCiO method can also minimize the sensor space. QCiO IoT network system accuracy is 93.25%, the precision of the system is 92.55%, and sensor sensitivity is 91.68% [30]. However, applying quantum technology in IoT platforms then it will improve its computational power and data processing capability.

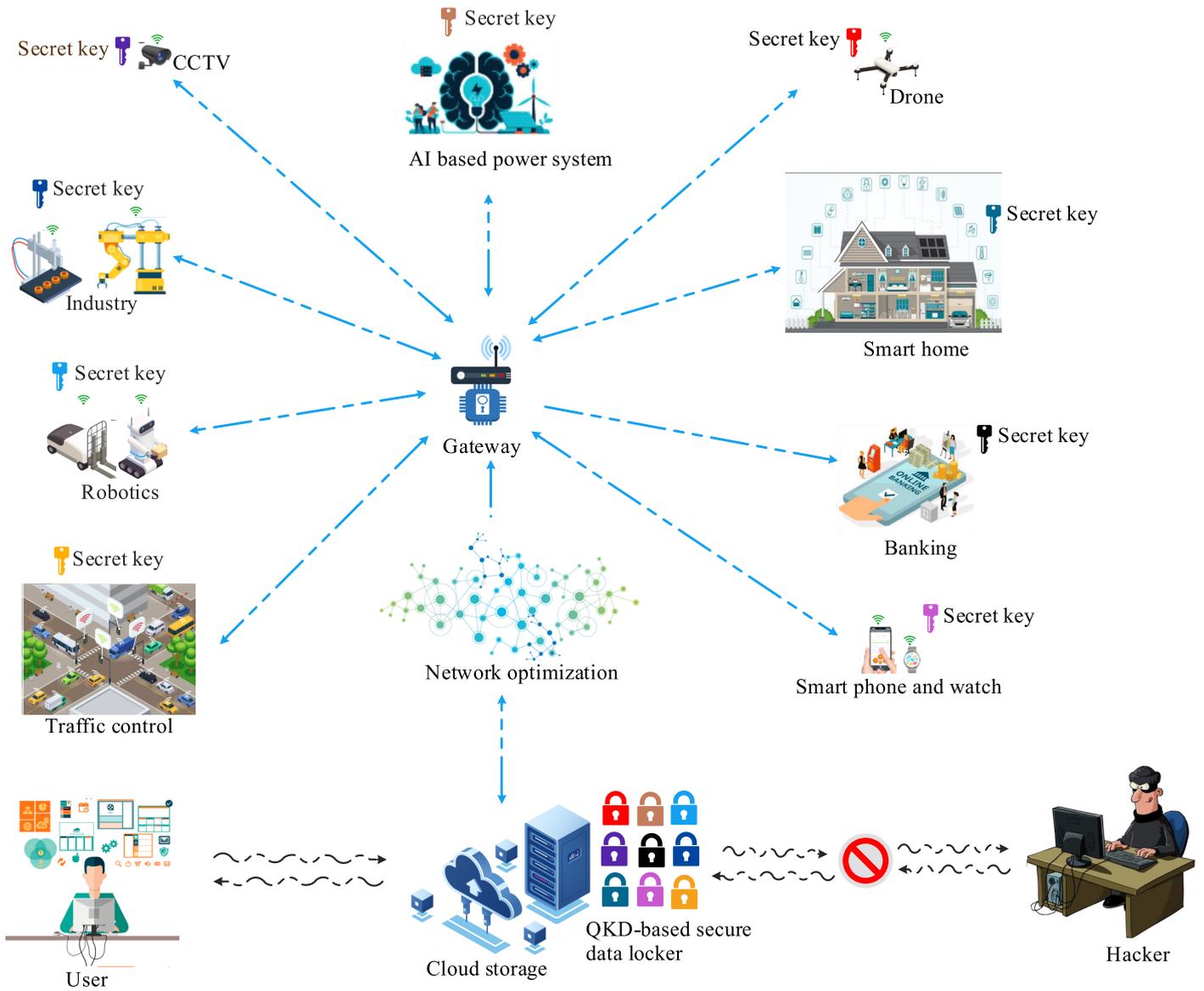

Fig. 11. IoT application for QKD based QCiO network system.

### 6.14 Quantum AI

Quantum mechanics has created a profound effect on the field of AI. The AI community strongly believes that quantum processing has great potential for solving some of the most difficult challenges we face now [94]. To solve the complicated problems and boost up the performance in different sectors such as data training, computing, pattern recognition, and communication quantum AI is a great approach [95]. Quantum AI works mainly in four steps. They are given below,

**Step 1:** Convert the quantum data to the quantum data set.

**Step 2:** Choose a quantum neural network model for quantum information processing in order to get proper information. That is hidden in an entangled state.

**Step 3:** Measurement of the quantum states collect classical information in the form of the samples from classical distribution. These values are obtained from quantum state.

**Step 4:** Evaluate classical neural network models for better data learning.

In quantum computation and communication, quantum AI creates a great impact. With the help of quantum AI applications it is now more easy to know the deeper structure of a human body, advancing the space exploration on a larger scale, create extensive connected security by combining the IoT and blockchain, and so on. Quantum AI is capable of reacting in both classical and quantum environment in any sensitive response [96].

### 7. Challenges and Future Research Directions

Several types of technical issues stand in the way of the implementation quantum communication network systems. These technological issues must be resolved before quantum communication systems can be properly deployed. Some of the major possible issues are briefly explained below:

i. **EPR Pair Supplement**: In quantum routing mechanisms and quantum transportation, EPR plays an essential role. When the quantum routing is executed, for each processing time EPR pair will consume. Quantum mobile devices and EPR pairs cannot be built up all the time because of wireless connectivity. For this reason, quantum wireless connections might fail to work by using EPR pairs. To solve this problem, there is require another charging device that can supply EPR pairs like quantum mobile station [16]. Their EPR can be said as an energy source.

ii. **Routing Path Exploration and Maintenance**: Initially, quantum wireless wide area network (WAN) was accepted for its network architecture. Quantum routing mechanism also works for both wireless local area networks and public area networks. To record the position of mobile devices, wireless WAN does not require network components like quantum bridge keeper (QBK) to transmit data wirelessly, because mobile devices shall explore routing paths automatically. Several routing path exploration techniques have been reported in the literature [16]. However, the majority of them are unsuitable for quantum-domain routing. The reason is conventional schemes give the privilege to find the routing path which is shortest. So, compared with conventional and quantum routing paths it shows that, the conventional routing path is based on a concise routing path, and quantum routing mechanism is based on EPR pairs. So, it's clear that the conventional routing path is still communicated in the routing path which is short. However, in the quantum routing mechanism if the numbers EPR pairs are short, then it will not be able to communicate [16]. As a result, if EPR pairs are insufficient then it will pick the latter path instead.

iii. **Handover Scheme Modification**: Taking care of EPR pairs is one of the main destinations of QBK. If any wireless quantum device shifts into another new coverage QBK's area, EPR pairs must have to archive share with the new QBK; make proper use of proposed EPR pair bridging, then this issue will be solved. Then the EPR generator is asked by QBK to set up pairs between one QBK to another QBK. After that, QBK creates the EPR-pair bridging to deliver the qubits, which are entangled with the qubits in the quantum mobile device, to QBK. Following that, the quantum mobile device would exchange EPR pairs with the QBK [16].

iv. **Error Rate:** Logic gates need to improve to achieve better performance. The intensity of laser light creates an environmental effect on qubits because laser produce both temperature and electromagnetic waves in nature which affect the qubits. About $10^{-6}$ error rates per gate can be eliminated by putting ions in miniature holes to prevent the unwanted transformation. In addition, error-correcting algorithms used in fault tolerance schemes may accept error probability rates of $10^{-6}$ per gate, which is sufficiently below the threshold accuracy [29].

v. **Decoherence:** Quantum computer's many problems are solved by quantum principle and its properties. These computers take advantage of quantum states. In future ICT, the next upcoming research challenge is decoherence. When qubits change their quantum states by interacting with the environment, it spoil informations in the way of destination. Decoherence is caused by a variety of factors, including radiation from heated objects, changing magnetic and electric fields, a collision between qubits, and the collapse of wave functions in the quantum mechanics [29]. That is why it is a challenging practical implementation issue for future ICT.

vi. **Barriers in Quantum AI:** Nowadays, all the researchers agree that the performance of quantum AI is better than conventional AI, but still this quantum AI technology faces some critical issues that create a barrier to its future progression. The training frameworks and open source data analysis are not available. Lacings of sustainable and well trained developer system environment. Insufficient amount of quantum hardware for converting classical data to quantum data. Scientists are relentlessly working on this quantum AI system to remove its barriers. Firstly, quantum algorithms must be used with practical quantum hardware in order to get its benefits. Secondly, in order to encode classical information in quantum mechanical form, quantum machine learning requires the integration of quantum random access memory as an interface device [97]. Finally, to achieve the full potential of quantum AI, applicability of quantum algorithms in quantum machine learning must be resolved. Researchers at Google company are working on a project on quantum algorithms to solve this issue [98].

vii. **Quantum State Fragility:** In the future, ICT quantum state fragility is one of the major challenges. We already know that quantum computer uses qubits for its computing. Compared with bits, qubits states are very weak using an environment for communication purposes. Quantum states might be modified for outdoor communication. This indicates that original pieces of information can likewise change or be lost in quantum communications in future ICT for outdoor conditions [29]. That is why quantum state fragility is a challenging issue for the future for uninterrupted connection and secure transmission.

viii. **Designing and Constructing Device:** The high performance network devices are so complicated to design and build because its programming language is not simple but rather very complex. Low attenuation or distortion can make a huge impact on the system. For this reason, transmitting the signal over a long distance with amplification is challenging.

ix. **Limited Resource:** New generation technologies come with a new infrastructure of the systems. Since quantum communication is a new technology, for this reason, all its resources are not available right now to complete the system infrastructure. Nowadays, scientists are in relentless pursuit of inventing new devices and instruments to overcome resource deficiency.

Limitations and disadvantages are the main challenges to overcome. To minimizes these limitations, hybrid technology and many other methods such as matter qubits and flying qubits are needed.

## 8. Conclusions

Over the past few decades, the applications of classical theories and principles have led the technological advancement in communication systems almost to its edge. Fascinating and innovative features are added to communication systems to overcome ongoing challenges. The way users are increasing, it can be easily predicted that even the technological development in 6G communication network systems cannot fulfill the growing demands and network security in the future. For this reason, to think outside of the box, scientists have introduced quantum physics, which has created a new path in modern physics. Scientists are giving hard and soul efforts to establish quantum based technological systems. Nowadays, several experiments are done to develop a shared infrastructure between quantum and classical communication systems. These technologies play a vital role in health care, space, banking, underwater communication, industry, and transportation. Besides, many other research and experiments are ongoing for the development of the quantum systems. In this review paper, we present definition and mechanism of quantum communication, the vision of quantum communication, difference between classical and quantum communication. That gives a clear overview of different types of implementation in quantum communication and its prospects. Quantum communication design goals and information processing have been demonstrated profoundly. We proposed quantum communication architecture. That clarifies the steps of quantum communication system from transmission to reception end, what will have to keep in mind before designing the system, and the information processing steps in quantum communication. Major QKD protocols functionality and their principles are explained elaborately. Numerous application scenarios are set out for different prospects of quantum technologies such as quantum internet, quantum computing, quantum based satellite communication, quantum underwater communication, and quantum IoT. Moreover, the challenges in quantum communication systems and future research directions are also illustrated. We believe that this review article will serve as a noble resource for comprehending the fundamentals, applications, protocols, and research directions of emerging quantum technologies.

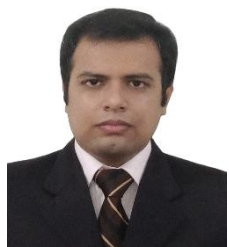

**Syed Rakib Hasan** received the B.Sc. degree in Electrical and Electronic Engineering from the Bangladesh Army International University of Science & Technology (BAIUST), Bangladesh, in May 2019, and he is currently pursuing M.Sc.Eng. degree in Electrical and Electronic Engineering from Khulna University of Engineering & Technology (KUET), Khulna, Bangladesh. His research interests include optical wireless communication, artificial intelligence, Internet of Things, reflective intelligent surface communication, wireless sensor networks, quantum communication, visible light communication, 5G and beyond communications.

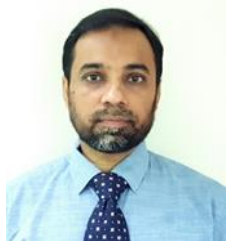

**Mostafa Zaman Chowdhury** received the B.Sc. degree in Electrical and Electronic Engineering from the Khulna University of Engineering & Technology (KUET), Bangladesh, in 2002, and the M.Sc. and Ph.D. degrees in Electronics Engineering from Kookmin University, South Korea, in 2008 and 2012, respectively. In 2003, he joined the Electrical and Electronic Engineering Department, KUET as a Lecturer, where he is currently working as a Professor. He has been working as a Head, Department of Biomedical Engineering, KUET since April 2022. He is the Senior Member of IEEE. He was a Postdoctoral Researcher with Kookmin University from 2017 to 2019 supported by National Research Foundation, Korea. He has published around 140 research papers in national and international conferences and journals. In 2008, he received the Excellent Student Award from Kookmin University. His three papers received the Best Paper Award at several international conferences around the world. He was involved in many Korean government projects. His research interests include convergence networks, QoS provisioning, small-cell networks, Internet of Things, eHealth, 5G and beyond communications, and optical wireless communication. He received the Best Reviewer Award 2018 by ICT Express journal. Moreover, he received the Education and Research Award 2018 given by Bangladesh Community in South Korea. He was a TPC Chair of the International Workshop on 5G/6G Mobile Communications in 2017 and 2018. He has been serving as Publicity Chair of the International Conference on Artificial Intelligence in Information and Communication (from 2019 to 2023) and International Conference on Ubiquitous and Future Networks ( 2022) . He served as a Reviewer for many international journals (including IEEE, Elsevier, Springer, ScienceDirect, MDPI, and Hindawi published journals) and IEEE conferences. He has been working as an Editor for ICT Express, an Associate Editor of IEEE ACCESS, an Associate Editor of Frontiers in Communications and Networks, a Lead Guest Editor for Wireless Communications and Mobile Computing, and a Guest Editor for Applied Sciences. He has served as a TPC member for many IEEE conferences.

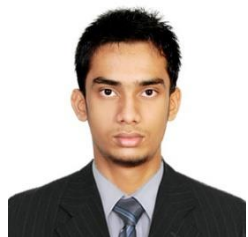

**Md. Saiam** received the B.Sc. degree in Electrical and Electronic Engineering from the Bangladesh Army International University of Science & Technology (BAIUST), Bangladesh in 2020. Currently, he is doing his M.Sc.Eng. degree in Electrical and Electronic Engineering at Khulna University of Engineering & Technology (KUET), Bangladesh. He is the Graduate Student Member of IEEE. He received awards from different national robotics and project competitions. His team was the champion of Project Showcasing in WUB 6th NATIONAL Com Tech Festival-2017, National Mechanical Festival IGNITION-2018 at KUET, and champion of Robo Soccer in BAIUST ROBOFEST-2018. He has successfully done several projects such as remote control aircraft design, android device controlled robot, line follower robot, gas leakage detection system, color object detection robot, water level detection system, RFID based security system, password-based door lock system, biometric door lock, biometric attendance system, fire alarm system, android-based home automation, smart car parking system, and modern traffic control system. His research interests include wireless communications and networking, wireless sensor networks, intelligent surface communication, Internet of Things, robotics and automation.

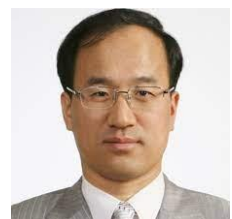

**YEONG MIN JANG** received the B.E. and M.E. degrees in electronics engineering from Kyungpook National University, Republic of Korea, in 1985 and 1987, respectively, and the Doctoral degree in computer science from the University of Massachusetts, USA, in 1999. From 1987 to 2000, he was with the Electronics and Telecommunications Research Institute (ETRI), Republic of Korea. Since 2002, he has been with the School of Electrical Engineering, Kookmin University, Seoul, Republic of Korea, where was the Director of the Ubiquitous IT Convergence Center from 2005 to 2010. He has been the Director of the LED Convergence Research Center, Kookmin University, since 2010, the Director of the Internet of Energy Research Center, Kookmin University, since 2018, and the Director of the AI Mobility Research Institute, Kookmin University, since 2021. His research interests include 5G/6G mobile communications, the Internet of Energy, the IoT platform, AI platform, eHealth, smart factory, optical wireless communications, optical camera communication, AI mobility, and the Internet of Things. Dr. Jang is also a fellow of the Korean Institute of Communications and Information Sciences (KICS). He had served as an Executive Director for KICS from 2006 to 2014. He was the President of KICS in 2019. He received the Young Scientist Award from the Korean Government from 2003 to 2006. He was a recipient of the Dr. Irwin Jacobs Award in 2018. He had also served as the Founding Chair for the KICS Technical Committee on Communication Networks in 2007 and 2008. He was/has been the Steering Chair of the Multi-Screen Service Forum from 2011 to 2019, the Society Safety System Forum from 2015 to 2021, and the ESG Convergence Forum since 2022. He has also served as the Chairman for the IEEE 802.15 Optical Camera Communications Study Group in 2014, and the IEEE 802.15.7m Optical Wireless Communications Task Group from 2015 to 2019 and successfully published IEEE 802.15.7-2018 and ISO 22738:2020 standard. He has been the Chairman of IEEE 802.15.7a Higher Rate and Longer Range OCC TG, since 2020. He has organized several conferences and workshops, such as the International Conference on Ubiquitous and Future Networks from 2009 to 2022, the International Conference on ICT Convergence from 2010 to 2022, the International Workshop on Optical Wireless LED Communication Networks from 2013 to 2016, the International Conference on Information Networking in 2015, and the International Conference on Artificial Intelligence in Information and Communication since 2019. He is also the Editor-in-Chief of ICT Express (indexed by SCIE). He has also been granted more than 120 patents.